# Methanofullerene Elongated Nanostructure Formation for Enhanced Organic Solar Cells

M. Reyes-Reyes[1*], R. López-Sandoval[2], J. Arenas-Alatorre[3], R. Garibay-Alonso[2], D. L. Carroll[4] and A. Lastras-Martinez[1]

[1]Instituto de Investigación en Comunicación Optica, Universidad Autónoma de San Luis Potosí, Alvaro Obregón 64, San Luis Potosí, Mexico.

[2]Instituto Potosino de Investigación Cientifíca y Tecnológica, Camino a la presa San José 2055, CP 78216. San Luis Potosí, Mexico.

[3]Instituto de Física, UNAM, Apartado Postal 20-364, 01000, México, D.F., Mexico

[4]Center for Nanotechnology and Molecular Materials, Department of Physics. Wake Forest University, Winston-Salem NC 27109, USA.

*Corresponding author. Instituto de Investigación en Comunicación Optica, Universidad Autónoma de San Luis Potosí, Alvaro Obregón 64, San Luis Potosí, Mexico. Tel.: +52-444-825-0183; fax: +52-444-825-0198.

E-mail address: reyesm@cactus.iico.uaslp.mx (M. Reyes-Reyes).

**Abstract**

Using transmission electron microscopy (TEM) and Z-contrast imaging we have demonstrated elongated nanostructure formation of fullerene derivative [6,6]-phenyl-C61-butyric acid methyl ester (PCBM) within an organic host through annealing. The annealing provides an enhanced mobility of the PCBM molecules and, with good initial dispersion, allows for the formation of exaggerated grain growth within the polymer host. We have assembled these nanostructures within the regioregular conjugated polymer poly(3-hexylthiophene) (P3HT). This PCBM elongated nanostructure formation maybe responsible for the very high efficiencies observed, at very low loadings of PCBM (1:0.6, polymer to PCBM), in annealed photovoltaics. Moreover, our high resolution TEM and electron energy loss spectroscopy studies clearly show that the PCBM crystals remain crystalline and are unaffected by the 200-keV electron beam.

**1. Introduction**

It is known that the polymer photovoltaics promise great flexibility, great processability and ever increasing efficiencies [1-3]. The fullerenes are efficient acceptor materials when they are blended with most conducting polymers. Yet, to realize efficiencies close to the theoretical maxima (set by the spectral overlap) in these bulk heterojunctions, internal resistance must be substantially lowered. The electron transport within the fullerene phase, i.e. between the interpenetrating network components, must be improved. Therefore, it is important to increase the mesoscopic order and the crystallinity of these interpenetrating networks. In the case of holes, this can be achieved with polymers undergoing crystallization with annealing, e.g. poly(3-hexylthiophene (P3HT). The P3HT [4] and fullerene derivative [6,6]-phenyl-C61-butyric acid methyl ester (PCBM) [5] crystallinities have been reported independently in blended film of PCBM:P3HT [4] and in PCBM film [5], respectively. Note that by combining both phases simultaneously, the charge balance can be maintained and in this way the internal resistance can be lowered by controlling the crystallinity in both phases simultaneously.

On the other hand, the enhanced crystallinity of the two components P3HT and PCBM in annealed blends has been also reported [2,6]. In a recent work [6] the morphology of PCBM:P3HT blend films has been studied by using bright-field (BF) transmission electron microscopy (TEM) images and selected-area electron diffraction (SAED) patterns. In that work different P3HT-PCBM concentration and solvent have been used than those we report here. Moreover, these results are related to low power conversion efficiency (2.7%) in comparison with previous results from our group [1,2] and W. Ma et al. [3], where 5% power conversion efficiencies were reported. We consider that the crystallinity of both materials is a very important feature as well as the morphology of PCBM crystals (i.e., agglomerates or elongate shaped PCBM nanocrystals). We think that the reported differences in efficiencies can be related to differences in the morphologies. The present report is focused to elucidate, using advanced techniques of electron microscopy, the morphology in films of high efficiency (5%) [2].

## 2. Experimental details

Photovoltaic devices were prepared by blending PCBM (American Dye Source) and P3HT (Aldrich: regio-regular Mw=87 kgmol$^{-1}$, without further purification) in chlorobenzene [1-2]. After that, the organic components were dissolved and filtered. The deposition of the solutions was carried out on glass/indium tin oxide (ITO, delta technologies Rs=20 Ohm-cm$^{-1}$) substrates. These substrates were thoroughly cleaned by using organic solvents and exposed to ozone for 90 min. The first film deposited by spin coating with a thickness of approximately 80 nm consist in poly(3,4-ethylenedioxythiophene)-poly(styrenesulfonate) (PEDOT:PSS, Baytron P). After drying this layer at 80 °C for 10 min, a second layer was deposited. The active layer was deposited through spin casting at 1500 rpm. These films were made with different P3HT:PCBM weight ratios 1:1, 1:0.6 and 1:0.4 corresponding to 15 mg of P3HT in 1 ml of chlorobenzene. The blend was continuously stirring for ~12 h.

The thickness of the blend layer for the P3HT:PCBM weight ratios 1:1, 1:0.6 and 1:0.4 corresponded approximately to 123 nm, 88 nm and 86 nm, respectively. Film thickness was measured by using a Tencor Alpha-step 500 surface profiler. The devices (without Al and LiF electrodes) were annealing in inert atmosphere (dry nitrogen glove box), after their cooling, they were prepared for an electron microscopy study. Note that the kinetic of the PCBM crystals can be reduced when annealing is performed on a real device with metal electrodes [3,6]. The samples for TEM measurements were prepared by floating the layers (PEDOT:PSS/P3HT:PCBM) onto a water surface, and transfer to a TEM grid. The photovoltaic characteristics for unannealed devices and for the optimally loaded annealed device have been previously reported [2].

BF TEM images and SAED patterns were recorded on a JEOL JEM-1230 transmission electron microscope operated at 100 kV. Z-contrast imaging, energy electron loss spectroscopy (EELS) and high-resolution transmission electron microscopy (HRTEM) were conducted on a JEOL 2010 FEG FAS transmission electron microscope operated at 200 kV (resolution 0.19 nm and spherical aberration 0.5 mm). X-ray diffraction (XRD) patterns were recorded using a Rigaku, DMAX 200 diffractometer using Cu K$\alpha$ radiation. For the study of the thin films in grazing incidence (GI) diffraction geometry [4], the angle between the film surface and the incident beam was fixed at 0.3°. The detector scans in a plane defined by the incident beam and the surface normal.

## 3. Results and discussion

Recently Reyes-Reyes et al. [2] have varied the loading of PCBM in P3HT films and have shown that the maximal efficiencies in P3HT:PCBM bulk heterojunction organic solar cells are found for very low PCBM loading (1:0.6). These devices with the optimal loading were annealed at 155ºC between two and three minutes resulting in an efficiency increase of up to 120% (from 2.4% to 5.2%). Several characteristics of the annealed and unannealed devices with different PCBM loading were analyzed to understand the great increase in their efficiency. In this paper, by using transmission electron microscopy, Z-contrast imaging and HRTEM, it is examined the morphology and meso-order of the PCBM nanostructures in these films.

First, we have observed that for all P3HT:PCBM weight ratios, 1:1, 1:0.6 and 1:0.4 corresponding to a PCBM weight percentage of 50, 37, 28, respectively, BF TEM images (not shown) show the morphology of the unannealed and annealed films to be quite homogeneous even at very high magnification (transmission electron microscope operated at 100 kV), similar results were presented by W. Ma et al. [3]. After that, we have performed a SAED analysis on unannealed films made at different concentration of PCBM (50, 37 and 28 wt %, Fig. 1a, b and c, respectively), three weak broad Debye-Scherrer diffraction rings can be distinguished with average d-spacings of 0.42, 0.25 and 0.13 nm. These same distances between adjacent planes are also observed in the annealed sample with optimal concentration (37 wt % of PCBM, Fig. 1d). There are several possible origins for the diffraction patterns in such films. However, it has recently been shown that these films are composed of homogeneously distributed PCBM nanocrystals [2]. Furthermore, through Near-Field Scanning Optical Microscope analysis, it has been proposed that these crystals take some kind of elongated shapes or large aggregations of PCBM [2].

In order to clarify the last assumption, a deeper and more elaborated study is necessary. The morphology of the films was observed by using techniques such as HRTEM,

Z-contrast imaging, EELS, and XRD. Z-contrast imaging provides a qualitative, atomic resolution, crystalline microanalysis in addition to show details at the interfaces between regions of different density [7]. The difference in densities between the P3HT (1.10 g cm$^{-3}$) [8] and PCBM (1.50 g cm$^{-3}$) [9] helps to discriminate between both components by using the Z-contrast procedure. Figures 1 and 2 show Z-contrast images at different loadings with and without annealing. Note that Z-contrast shows the phase segregation between the polymer and PCBM in the films. With HRTEM and EELS we have corroborated, in the Z-contrast images, that the PCBM agglomerates correspond to the brighter areas of the figures and, as a consequence, the rest of the material, the darker areas, is the polymer.

The effect of concentration and annealing is evident when the samples are observed by Z-contrast imaging. Figure 1 shows results obtained from a device at 1:1 weight ratio (P3HT:PCBM) unannealed (Fig. 1e and 1f) and annealed at 155°C by 3 min (Fig. 1g and 1h). From the Z-contrast image of the sample before annealing (Fig. 1e) we can observe clearly that the film is composed of many PCBM agglomerates with sizes around 300 nm distributed homogeneously and some others of approximately 0.5 μm. On the other hand, the annealed film displays extensive flat areas with some large PCBM agglomerates (around 1.2 μm in length and 0.5 μm in width) as seen in Fig. 1g. One reason for finding this kind of morphology is because the agglomerates, formed before annealing, begin to merge during the annealing time.

Long crystallites were difficult to observe in our conventional TEM images, even SAED patterns, too weak, were not an enough proof for showing their presence. However, Z-contrast images, HRTEM and EELS indicated that small "crystalline" pathways are well dispersed throughout the polymer matrix. Yang et al. [10] suggested this in a different system that the reported in this work; they observed the formation of PCBM single crystals by using SAED patterns. We have corroborated the existence of these PCBM nanocrystals using HRTEM. HRTEM image of the unannealed film at 1:1 weight percentage (Fig. 1f) shows some atomic planes of the unannealed sample; some crystallization occurs for the PCBM because there is a high amount of this material. These unannealed films present nanocrystals with different sizes and the d-spacings correspond to 0.25 nm (region 1 in Fig. 1f) and 0.29 nm (region 2, and 3 in Fig. 1f).

On the other hand, our results show that unannealed films at low concentration do not exhibit agglomerates of PCBM (see Fig. 2a). Figure 2 shows images of films made at 1:0.6 weight ratio of P3HT:PCBM. The Z-contrast image reveals "flat" areas (Fig. 2a), not agglomerated of PCBM, and t no crystalline planes were observed by HRTEM (Fig. 2b). Nevertheless, SAED patterns show some rings (see Fig. 1c) and therefore we can not totally eliminate the possibility of their presence since we only observe the surface of the films. Anyway, they are relatively small in number. When these films are annealed however, we can observe in Z-contrast images a surprising change in the PCBM morphology (Fig. 2c). Figure 2c shows PCBM agglomerates, as elongated nanostructures (brigther areas, pointed out by arrows). These elongated nanostructures are homogeneously distributed across the surface of the film. We propose that these nanostructures extend from the inside part of the film. Notice that we are only observing by TEM one of the surface of the film. The approximated length of the PCBM elongated nanostructures ranges from 35 nm to 135 nm and width around 15 nm (on the surface). HRTEM images show elongated nanostructures composed by several nanocrystals arranged at different orientations, and with d-spacings of 0.42, 0.24. 0.25 and 0.24 nm shown in 1, 2, 3 and 4 regions, respectively (Fig. 2d arrows show one of these structures). Values near of these PCBM d-spacings have been reported [5]. In the structure shown in Figure 2d, one can notice that some areas are not very clear because the nanodomains are covered with the polymer. Regarding the P3HT, it has been reported by several groups that annealing increases their crystallinity [3,4]. The effects of the annealing time on the films with low PCBM concentration are shown in Figs. 2e-f. The last image (Fig. 2f) is an amplification of the inset in Figure 2e. In these cases we show the films at the optimal loading (1:0.6/P3HT:PCBM) annealed at 120°C during 60 min. Notice the change in the morphology and the size of these PCBM nanostructures in comparison with the case of short annealing time at 155°C (Fig. 2c,d). In addition, these morphologies are different with those reported to high concentration of PCBM in PCBM:P3HT blended film [11]. These images show clearly that it is very important to consider the morphology of PCBM agglomerates in order to enhance the efficiency in these systems. Finally, there are differences in brightness in certain regions of the PCBM structures (Figs 2e-f) that are due to the depth: these differences depend if these regions are on the surface or in the bulk of the film. This proves that these PCBM nanostructures come from the inside part of the film to the surface. This result was bore out using bright field transmission electron microscopy (not shown).

HRTEM images clearly resolve, in real space, the structure of PCBM nanocrystals. The inset of Figure 2f shows a magnified HRTEM image of PCBM nanocrystal in P3HT host after annealing at 120 $^0$C for 60 min at optimal loading (1:0.6/P3HT:PCBM). The same HRTEM images have been found for different PCBM loadings and annealing times. We use this particular image only as example. The d-spacing in this PCBM crystal is about 0.42 nm.

Figure 3a shows the EELS spectrum taken from a PCBM crystal. This crystal was first observed by Z contrast and after that we got the EELS spectrum. In the inset of Figure 3a, we show the characteristic 284-eV K-edge related to the excitation to the state $\pi^*$ of electrons from the K-shell, which is a signature of the existence of $sp^2$ bond in the material and it is found in the graphite, carbon nanotubes and $C_{60}$ structures. We also note the remarkable stability of the PCBM nanostructure subjected to a 200-keV electron beam during both the HRTEM and EELS studies (about 30 min). However, the structural stability of PCBM nanostructures is in sharp contrast to the behavior of pristine $C_{60}$ structures however [12-14].

Figures 3b, c and d show a X-ray diffractogram of the annealed P3HT/PCBM films at different concentrations (50, 37 and 28 wt %, respectively) on glass/ITO/PEDOT-PSS substrate. From the X-ray patterns of the annealed sample and comparing with previous results [4,15-17], the detected peak at $2\theta = 5.36°$ arises from polymer crystallites with a lattice constant about $1.64\pm0.2$ nm and a-axis orientation. In this way, the main chain is parallel to the substrate whereas the side chains are perpendicular to it.

Crystalline P3HT domains have been observed after an annealing treatment [4]. The effect on P3HT crystallization by annealing at different PCBM concentrations was also studied using XRD. Figure 3d shows that the size of (100) reflections increases when we decrease the concentration of PCBM. This suggests that the low weight percentage of PCBM permits better crystallization of the P3HT polymeric chains. And, comparing with other results [4,8,15-16], our data indicate that the main chains of the polymer have a tendency to orient themselves parallel to the film surface, no crystallites with b or c axis orientation were detected (Fig. 3b, c, d). On the other hand, the narrow peaks likely corresponding to PCBM were observed with lattice spacing of 0.42, 0.29, 0.25 and 0.24 nm. These values are consistent with those determined from the HRTEM images. However, Ma et al. [3] suggest that the peaks at $2\theta \approx 13°$ (d $\approx$ 0.82 nm) arise from the PEDOT layer and the peaks at $2\theta \approx 22°$ (d $\approx$ 0.42 nm) and 31° (d $\approx$ 0.29 nm) are due to ITO. They also show a X-ray peak at $2\theta \approx 35°$ (d $\approx$ 0.25 nm) upon annealing, but this result is not discussed. However, contrary to the observations by Ma et al. [3], Erb et al. [4], and Zhokhavets et al. [17], who did not find diffraction peaks corresponding to the PCBM in the P3HT/PCBM films, we have found them using Z-contrast imaging and HRTEM crystals of PCBM molecules. In addition, the peak at $2\theta \approx 35°$ is a strong evidence for PCBM nanocrystals, where we have found the same X-ray peak in PCBM pristine films (not shown). We also show that when we increase the PCBM loading or the annealing time the sizes of these PCBM structures are enhanced. However, these PCBM structures are composed of a set of PCBM nanocrystals. Possible reasons for the differences shown above could be related to the solvents [4] and the experimental processes [17] used for preparing the active layer and/or annealing temperature. Another possibility is that these peaks are overlapped with peaks from the ITO and PEDOT film [3]. In order to minimize the reflections caused by the ITO and PEDOT film, the annealed film at 37 wt % of PCBM was studied in GI diffraction geometry (Figure 3e). Figure 3e presents the comparison between XRD and GI XRD. GI XRD measurements show that the peaks correspond to PCBM crystals and they are in agreement with HRTEM results (0.42, 0.29, 0.25 and 0.24 nm). Therefore, our results suggest that the PCBM X-ray peaks are overlapped with ITO and PEDOT X-ray peaks hindering their identification using conventional XRD.

## 4. Conclusion

Our experimental results show that the formation and morphology of PCBM elongated nanostructures can be controlled using, annealing temperature, annealing time, and the PCBM concentration. Further, the formation of the crystalline PCBM elongated nanostructures favors polymer crystallization without inducing the abrupt phase segregation that is, in general, very unfavorable for high efficiency photovoltaic devices. We show that these elongated nanostructures are composed of several nanocrystals with different orientations. In addition, our HRTEM and EELS studies clearly show that the PCBM crystals remain crystalline and are unaffected by the 200-keV electron beam.

**Acknowledgments**

We want to thank L. Rendón (IF-UNAM), I. Esparza-Alvarez (IM-UASLP) for the technical assistance and facilitating the use of the microscope based in Mexico and L. Narváez (IM-UASLP) for the XRD measurements. This work was supported at UASLP by FAI through grant no. C05-FAI-10-17.38, PIFI grant no. C06-PIFI-03.6.6 and by CONACYT 2005-J48897-Y. At UNAM by CONACYT 2003-C02-4450 and Wake Forest University by the Air Force Office of Sponsored Research (AFOSR) under grant no. FA9550-04-1-0161.

**Figure Captions**

Fig. 1. SAED patterns of P3HT:PCBM unannealed (a,b,c) and annealed at 155 °C for 3 min (d) films prepared with PCBM concentrations of 50 (a), 37 (b,d) and 28 wt % (c). Z-contrast (e,g) and HRTEM (f,h) images of P3HT:PCBM films of 50 wt % before (e,f) and after annealing (g,h). The inset in (h) shows a TEM image of a PCBM nanocrystal. The numbers inside (f), and (h) show crystallized domains.

Fig. 2. Z-contrast (a,c) and HRTEM images (b,d) of PCBM:P3HT films with 37 wt % of PCBM before (a,b) and (c,d) after annealing (155 °C, 3 min). Elongated nanostructures are shown by arrows in (c) and (d). The numbers inside (d) show crystallized domains. The effect of the annealing time (120 °C, 60 min) in the film at low PCBM concentration is shown in (e-f). The inset in (f), we show a high-resolution transmission electron micrograph of PCBM (37 wt %) showing the chain-like structure of this material.

Fig. 3. EELS spectrum (a) taken from a PCBM crystal. Diffractogram of PCBM:P3HT composite films deposited on glass/ITO/PEDOT:PSS prepared with PCBM concentrations of 50 (b), 37 (c) and, 28 (d) wt %. Comparison between XRD and GI XRD with 37 wt % of PCBM is shown in (e). This figure shows that our film continues presenting peaks at $2\theta \approx$ 22°, 31° and 35°.

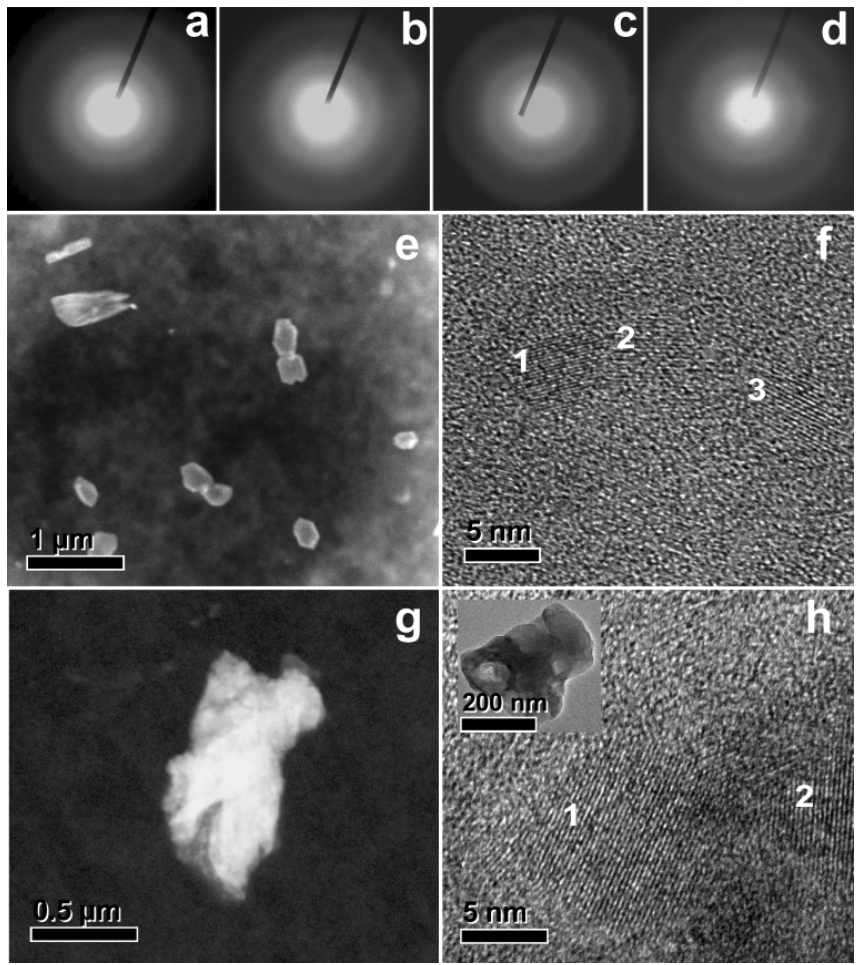

**Figure 1. Reyes-Reyes M. et al.**

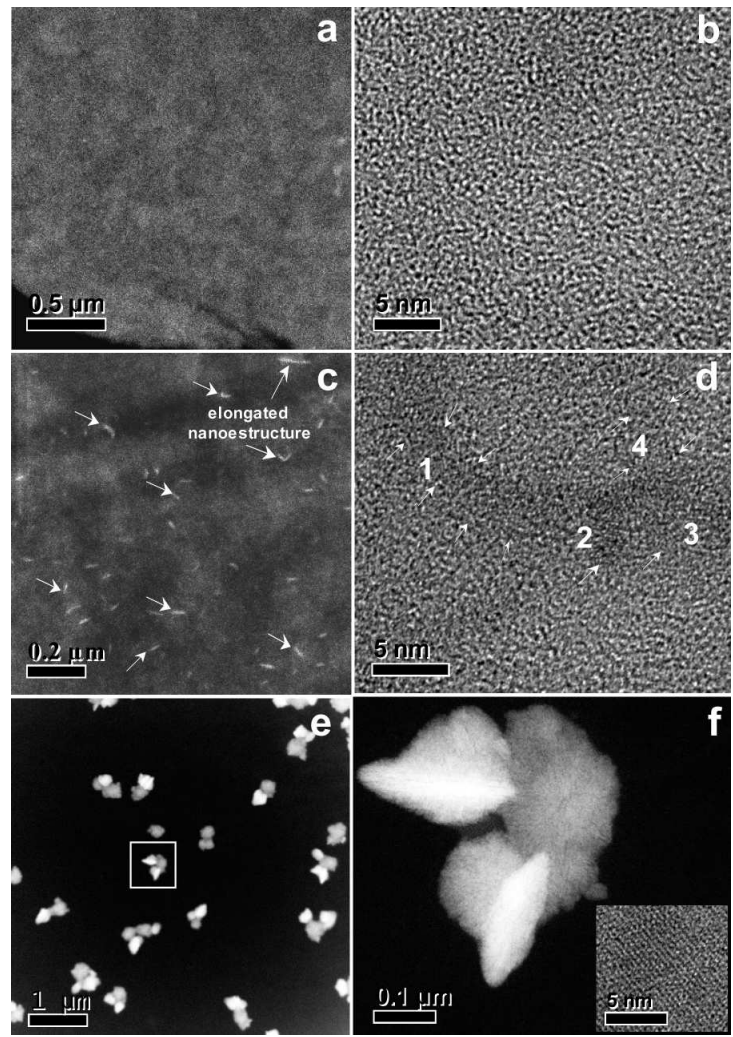

**Figure 2. Reyes-Reyes M. et al.**

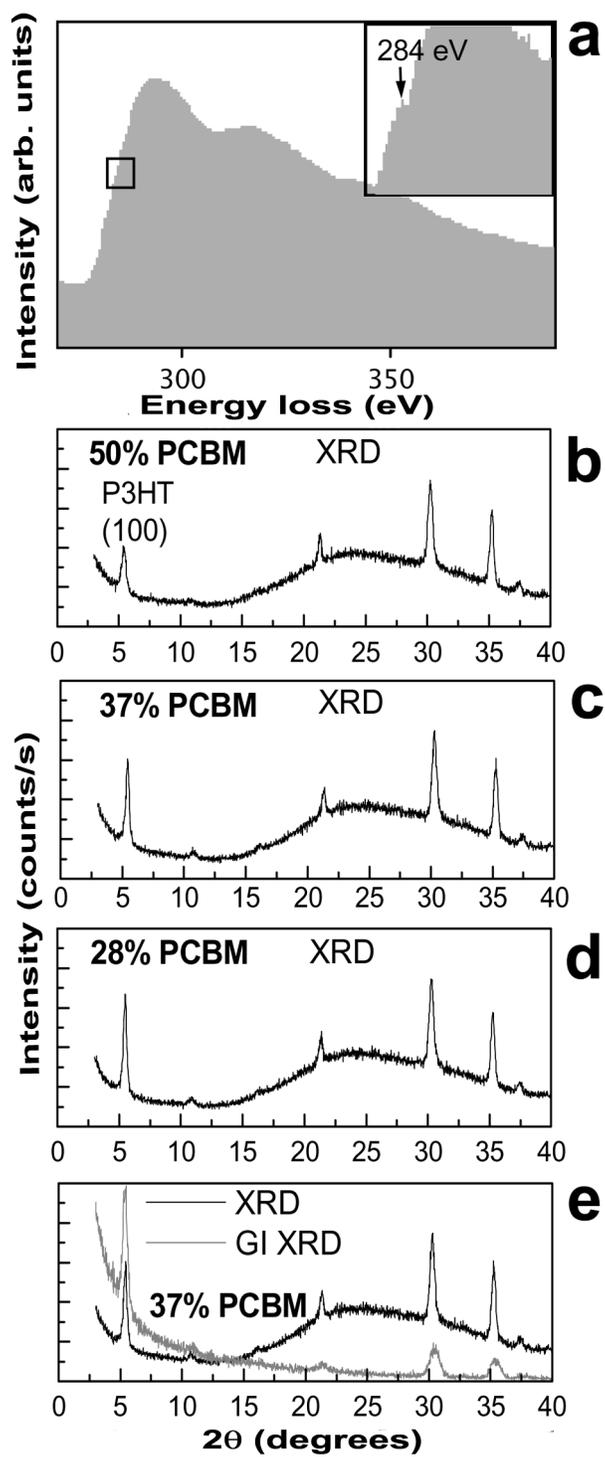

**Figure 3. Reyes-Reyes M. et al.**